\documentclass[12pt,journal]{IEEEtran}
\usepackage{colortbl}
\usepackage[pdftex]{graphicx}
\usepackage{color, colortbl}
\definecolor{Row0Color}{rgb}{0.83, 0.83, 0.83}
\definecolor{Row1Color}{rgb}{0.95, 0.95, 0.96}
\definecolor{Row2Color}{rgb}{0.83, 0.83, 0.83}
\usepackage{balance,soul}
\usepackage{color}
\usepackage[table]{xcolor}
\usepackage{siunitx}
\usepackage{url}
\usepackage{marginnote}
\usepackage{booktabs}
\usepackage{multirow}
\usepackage{cite}
\usepackage[T1]{fontenc}
\usepackage{todonotes}
\usepackage{subcaption}
\usepackage{float}
\usepackage{svg}
\newcommand{\modi}[1]{{\textcolor{black}{#1}}}

\begin{document}

%\title{Approaches and challenges in next generation GNSS interference management: leveraging connectivity and data}
% \title{Trends and challenges in next generation GNSS interference management}
%\title{Leveraging Data and Connectivity in Next Generation GNSS Interference Management}
%\title{Data-driven Interference Management for the Next Generation GNSS receivers}
\title{Trends and Challenges in Next-Generation GNSS Interference Management}
% \title{Approaches and challenges in reinforcement learning and factor graph optimization for GNSS interference management}

% Petri: I added my name, but I am not sure about the order, rearrange if necessary
%\author{Leatile Marata,~\IEEEmembership{Member,~IEEE,}
  %      Mariona Jaramillo-Civill,~\IEEEmembership{Member,~IEEE,}
    %    Tales Imbiriba,~\IEEEmembership{Member,~IEEE,}
     %   Petri Välisuo,~\IEEEmembership{Member,~IEEE,}
     %   Heidi Kuusniemi,~\IEEEmembership{Member,~IEEE,}
      %  Elena Simona Lohan,~\IEEEmembership{Senior Member,~IEEE,}
%and~Pau Closas~\IEEEmembership{Senior Member,~IEEE}
        % <-this % stops a space
%\thanks{E.S. Lohan and R. Morales Ferre are with Tampere University. H. Kuusniemi and M. Elsanhoury are with University of Vaasa. J. Praks is with Aalto University. S. Kaasalainen and C. Pinell are with the Finnish Geospatial Research Institute.}% <-this % stops a space

%}% <-this % stops a space
\author{\IEEEauthorblockN{Leatile Marata, \textit{Member, IEEE}, Mariona Jaramillo-Civill,~\IEEEmembership{Member,~IEEE,} Tales Imbiriba,~\IEEEmembership{Member,~IEEE,} Petri Välisuo,~\IEEEmembership{Member,~IEEE,}Heidi Kuusniemi,~\IEEEmembership{Member,~IEEE,}
        Elena Simona Lohan,~\IEEEmembership{Senior Member,~IEEE,} and Pau Closas~\IEEEmembership{Senior Member,~IEEE}}
\thanks{Leatile Marata and Elena Simona Lohan are with Tampere Wireless Research Centre, FI-33014, Tampere University, Finland (e-mail: leatile.marata@tuni.fi, elena-simona.lohan@tuni.fi). Mariona Jaramillo-Civill and Pau Closas are with the Department of Electrical and Computer Engineering, Northeastern University, Boston, MA, USA (e-mail:jaramillocivill.m@northeastern.edu, closas@northeastern.edu). Tales Imbiriba is with the Department of Computer Science, University of Massachusetts Boston (UMB), Boston, MA, USA (e-mail: Tales.Imbiriba@umb.edu). Petri Välisuo and Heidi Kuusniemi are with the University of Vaasa, Finland (e-mail: petri.valisuo@uwasa.fi, heidi.kuusniemi@uwasa.fi).}}

\markboth{AESM-2025-0077}%
{LastName \MakeLowercase{\textit{et al.}}: }
%\markboth{IEEE Communications Magazine}%
%{LastName \MakeLowercase{\textit{et al.}}: Leveraging data and Connectivity in next generation GNSS interference management: Approaches and challenges}

\maketitle

\begin{abstract}
The global navigation satellite system (GNSS) continues to evolve in order to meet the demands of emerging  applications such as autonomous driving and smart environmental monitoring. However, these advancements are accompanied by a rise in interference threats, which can significantly compromise the reliability and safety of GNSS. Such interference \modi{problems} are \modi{typically} addressed through signal-processing techniques that rely on physics-based mathematical models. Unfortunately, solutions of this nature can \modi{often} fail to fully capture \modi{the} complex forms of interference. To address this, artificial intelligence (AI)-inspired solutions are expected to play a key role in future interference management solutions, thanks to their ability to exploit data in addition to physics-based models. This \modi{magazine paper  discusses the main challenges and tasks required to secure GNSS and present a research vision on how AI can be leveraged towards achieving more robust GNSS-based positioning}.
%\hl{The global navigation satellite system (GNSS) continues to evolve in order to meet the demands of emerging  applications such as autonomous driving and smart environmental monitoring. However, these advancements are accompanied by a rise in unintentional and intentional interference threats, which can significantly compromise the reliability and safety of GNSS. Such interference is normally addressed through signal-processing techniques that rely on physics-inspired mathematical models. Such solutions can sometimes fail to fully capture the complexity of different forms of interference. Owing to their ability to exploit data in complex scenarios, artificial-intelligence inspired solutions are expected to come in handy in future interference management solutions and, consequently, enhance the robustness of GNSS. The contribution of this article to the body of knowledge on this advancement is twofold: (i) to discus the main challenges in GNSS interference management, namely classification, detection, localization, and mitigation; and (ii) to present a research outlook on how AI can be leveraged for GNSS interference management.}

% THIS IS HOW WE HAD INITIALLY WRITTEN 
%\textcolor{red}{In this article, we discuss the main challenges in GNSS interference management, namely classification, detection, localization, and mitigation, as well as the research outlook in light of advances in various fields.}

\end{abstract}

%\IEEEpeerreviewmaketitle

\section{Introduction}
\label{label_Introduction}
%\pau{introduce PNT in general, then move to satellite-based PNT}
%\hl{Please note that we can have maximum 15  references in a magazine paper; we already have more than 15}
%\LM{I tried to improve the flow in the intro.}
\par The need for a digitally connected society has led to the emergence of novel technological applications, such as smart environmental monitoring, smart farming, and autonomous vehicles \cite{Prol2022}. These applications require precise positioning, navigation, and timing (PNT) parameters  to function \cite{Celikbilek2022,reid2020,kassas2020}. As such, Global Navigation Satellite Systems (GNSS) is increasingly playing a crucial role in providing such PNT solutions. In fact, GNSS has been widely used in safety-critical applications such as terrestrial, maritime, and aerial transportation \cite{kassas2020}. Owing to  increasing demand in robust and accurate positioning solutions, there is a renewed research interest aimed at improving traditional GNSS solutions that relied on Medium Earth Orbit (MEO) satellites. For instance, new paradigms that rely on Low Earth Orbit (LEO) satellites are emerging, thus further enhancing the GNSS coverage \cite{Prol2022,Celikbilek2022,reid2020,kassas2020}. Unfortunately, this proliferation of GNSS applications has also triggered novel interference threats that can render GNSS unreliable and prone to risks \cite{Dovis15,Ioannides2016PIEEE,borio2016impact}. Therefore, there is a need to address such vulnerabilities in a more comprehensive manner. It is equally important to note that such interferences are also getting more advanced or sophisticated. This is due to a huge availability of cheap resources, such, as software defined radios (SDR) that can  be used to generate GNSS interferences \cite{curran2016threat}. 
\par Recent research has revealed a significant increase in cyber-attacks in the L- and S- satellite bands in the past few years \cite{Ferre2020,Korobko2021}. For instance, Korobko \textit{et al.} in \cite{Korobko2021} reported a malicious positioning system interference that led to the emergency landing of a long-range unmanned autonomous vehicle (UAV) due to Global Position System (GPS) jamming in Ukraine. In the same light, authors in \cite{Staff22} reported some disruptions in GNSS services that hindered the commercial air traffic in central and eastern Finland. Consistent with these, the GPSJAM\footnote{www.gpsjam.org} webpage constantly tracks daily maps of GPS interference and indeed they are consistently increasing, further indicating a high likelihood of continued growth.  In GNSS, such attacks can either be unintentional or intentional. The former arises from either channel impairments, such as multipath or signal attenuation due to weather conditions, or existing communication and navigation systems that transmit (legally) on nearby or the same frequency bands as GNSS do. On the other side, intentional interferences stem from well-crafted artificial security threats, which can be further classified into jammers and spoofers. A jammer signal is a high power interference that seeks to overpower the genuine GNSS signal and lead to its complete loss, while spoofers are generated as counterfeit GNSS signals, intended to deceive receivers. A sub-category of spoofers are the meaconers, where a delayed version of a genuine GNSS signal is re-transmitted with the intention of misleading the receiver about its position and time \cite{zidan2020gnss,Ferre2020,Korobko2021}.\par 
As noted in the preceeding paragraph, the originators of intentional interference have clear-cut motives and thus dealing with such interference requires more advanced interventions to avert security threats. Unfortunately, as the number and types of interference threats increases, conventional threats detection mechanisms are becoming less efficient \cite{10379436}. Notably, conventional interference management techniques rely on advanced signal-processing techniques that are based on specific mathematical models of the interference. As such, these models cannot fully capture complex interference scenarios such as co-existing multiple types of jamming and spoofing \cite{Ferre2020}. Alternatively, encryption-based techniques can be considered for problems such as spoofing detection \cite{10706977}. However, these have not been fully developed and are not yet widespread. All in all, existing interference management solutions for GNSS receivers are limited by their lack of adaptability to emerging and novel security threats, which can pose significant risks to GNSS\footnote{\modi{We would like to clarify that, although GNSS interference management faces significantly broader challenges including data management with interfaces and rights management across different authorities, as well as regulatory measures (legal/technical) our focus here is specifically on the technical aspects.}}. Recently, there has been a growing interest in Artificial Intelligence (AI) driven interference management techniques, due to their ability to learn from available interference data and enhance their performance over time. This adaptability makes AI based approaches particularly promising for addressing previously unseen threats or complex-to-model effects. This \modi{magazine paper} aims to serve as a resource for researchers interested in developing AI-enhanced GNSS solutions. Specifically, we discuss AI-inspired interference management strategies that leverage massive data from users sharing an area, such as a smart city, to detect, classify, localize, and mitigate GNSS interferences.  \modi{We would like to emphasize that this is a magazine paper and, as such, it does not aim to propose novel technical solution but rather to highlight new potential areas of research on GNSS interference using AI. The summarized contributions are as follows:}

\begin{itemize}
    \item We discuss the challenges and the state-of-the-art solutions in interference management for GNSS receivers. Specifically, we focus on existing signal-processing mechanisms that rely on mathematical assumptions and reveal their successes and limitations in meeting the current demands of location accuracy and robustness.  
    
    \item  We provide a review of current and future directions in the fast evolving field of GNSS interference detection, classification, localization, and mitigation. 
    
     \item \modi{We discuss} how AI can be leveraged, \modi{in novel manners} for interference management \modi{as well as} the relevance of \modi{AI-based} distributed architectures. \modi{In addition, we discuss} the benefits of hybridizing GNSS with alternative systems and sensors, \modi{and} the need to develop techniques that can adapt to new interference threats. \modi{For instance, we discuss how the machine learning (ML) solutions could exploit the correlations between features of GNSS signals at different stages of the receiver, such as pre-correlation and post-correlation, in order to improve the threat detection accuracy.}  %For instance, we envision the joint detection and classification of interference, which can facilitate the receiver to adaptively choose the most suitable interference rejection scheme.  
      % \item We also propose novel machine learning (ML) solutions that exploit the correlations between features of GNSS signals at different stages of the receiver, such as pre-correlation and post-correlation, thereby improving threat detection accuracy.  
  
    % \item We also propose a distributed data fusion technique that generates global interference probability maps from local threat probability maps over space. This in turn can serve as prior information for correcting navigational data, for instance, when using factor graphs.
  %  \item The article presents preliminary, real-world results in the context of Jammertest 2024 experiments, where the impact of both jamming and spoofing to commercial receivers is discussed.
  \item \modi{Last but not least, we present preliminary real-world results from the Jammertest 2024 experiments, where various live jamming and spoofing signals were studied by different stakeholders, including research institutes and companies. While we do not aim to provide a full technical analysis of these results, they serve as a reminder of existing benchmark rich datasets with real-field wireless interferences.} 
\end{itemize}

The remainder of this article is organized as follows. Section \ref{sec:int_manag} discusses the main challenges and recent solutions in GNSS interference detection, classification, localization, and mitigation. A research outlook is also provided in Section \ref{sec:vision}, pointing out emerging topics and trends in this rapidly evolving field.
The effects of different types of interferences are demonstrated with over-the-air experiments in the context of the Jammertest 2024 campaign, whose preliminary results are discussed in Section \ref{Sec:jammertest}. Section \ref{sec:conclusions} concludes the article with final remarks.
    
%################
% \section{Interference threats}\label{sec:int_threats}

%\pau{quickly review what are the main interference types, i.e. jamming and spoofing, and their effects. Has been said many times, so can be a short section.}
%\pau{reiterate the need for interference monitoring: critical infrastructures, safety-critical applications, location-awareness in autonomous systems, etc.}

% As mentioned earlier, interference degrades the performance of a GNSS system. Specifically, jammers block the signal, thus, preventing the GNSS signal from being received.  The spoofer interference is characterized by creating a fake version of the original GNSS signal, with the intention to mislead the GNSS receiver compute wrong PTN. As such, the spoofer impact is more pronounced than the jammer impact as the GNSS as systems that rely on GNSS, e.g., robots, aircraft, are mislead to navigate using wrong routes, thus, leading to accidents. It is therefore crucial, especially in critical infrastructures such as safety applications, to manage such interference based on its nature \cite{Ferre2020}. 
%\pau{maybe a summary table?}
%\hl{SL: yes, we need a summary table or block diagram  with different interference types }
%################
\section{State-of-the-art in Interference Management}\label{sec:int_manag}

%\pau{expand on the ongoing efforts and future directions for each of the 4 elements in interference management}
%\pau{too short of a section. I suggest adding more refs and details about the 4 elements in interference management. Kind of a summary of \cite{Ferre2020}} 
%\LM{NOTED WITH THANKS}
%\pau{maybe consider a receiver diagram where the blocks in Table 1 can be identified.}

\begin{figure*}[t!]
    \centering    \includegraphics[width=\textwidth]{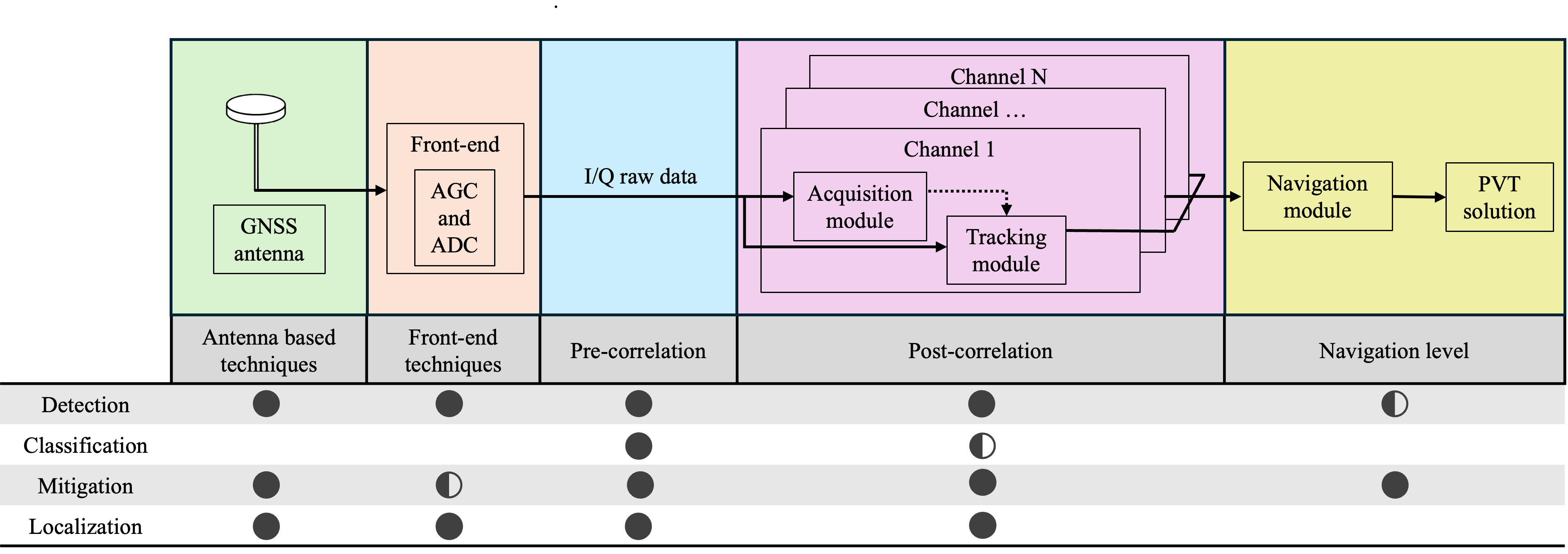} 
    \caption{GNSS receiver stages and typical places for interference detection, classification, localization, and mitigation. The table at the bottom uses filled circles to indicate primary implementation locations and half-filled circles to represent secondary or less common implementation locations for each task across different receiver stages.}

    \label{fig:rxstages}
\end{figure*}

 %\hl{In order to strengthen GNSS signal integrity and mitigate intentional interference threats, various solutions have been developed in the literature. These solutions can be classified according to their functionality: detection, classification, localization, or mitigation and they are illustrated in Fig. \ref{fig:rxstages} in the context of the receiver blocks of a typical GNSS receiver.}%% 

 In order to strengthen GNSS signal integrity and mitigate intentional interference threats, various solutions have been developed in the literature. These solutions can be classified according to their functionality: detection, classification, localization, or mitigation as illustrated in Fig. \ref{fig:rxstages} of typical blocks of a GNSS receiver. In the figure, the filled and half-filled circles in the lower table indicate the relationship between receiver stages and interference management functions. Fully filled circles suggest the typical location of the corresponding tasks, while half-filled circles illustrate possible theoretical locations, which are not much used in the current literature. The \textit{detection} block comprise of techniques that determine the presence or absence of interference, while  \textit{localization} is concerned with identifying the exact position of interference signal \cite{Ferre2020}. On the other hand, the \textit{mitigation} block is made of techniques that reduce (where possible eliminating) the impact of interference, while \textit{classification} algorithms determine the type of interference, i.e., whether it is a jammer or a spoofing signal. Furthermore, such solutions can also be classified according to the stages at which they are applied in the receiver \cite{8060594}. We present a summary of these in Table~\ref{signalProcessingTable}, where the capital letter acronyms  D, C, L, and M in brackets in the first column stand for detection, classification, localization, and mitigation, respectively. It can be noted that most existing solutions are limited in their inability to handle complex interference and are also prone to high complexity processing requirements. Equally important to note is that \modi{the} majority of solutions in Table~\ref{signalProcessingTable} operate independently at different receiver stages. As such, they fail to exploit the correlations that could reveal relationships between the structures of the signal at different stages of the receiver, which has potential to enhance interference management \modi{according to our vision and preliminary findings}. Next, based on current literature and the community trend, we present our vision for interference management aimed at addressing the outlined shortcomings.

%\hl{In Table~\ref{signalProcessingTable}, the capital letters in brackets in the first column are the acronyms D, C, L, M for detection, classification, localization, and mitigation, respectively. It can be noted that most of the existing solution are limited due to their inability to deal with complex interferes. Moreover, it can be noted that most existing solutions  from Table~\ref{signalProcessingTable} work at different receiver stages, independent of the other stages. Next, we present our vision for interference management aimed at addressing the outlined shortcomings. }

%\textcolor{red}{A a signal processing solution, detection can be done at automatic gain control, digital precorelation or post corelation. Some works aimed at this include \cite{7444136}, where  Gao \textit{et al} propose spatial filtering and time-frequency vector tracking for detection of jammers. On the other hand, beamforming based techniques such as \cite{7444122} can also be exploited to increase the robustness of GNSS receiver. On the other hand, some data encryption methods have also been of interest. One of these include the recent efforts in using cryptographic defenses have been developed Galileo GNSS
%European system Open Service Navigation Message Authentication (OSNMA)}

%\pau{might be good to have a summary/classification table for these elements}
%\LM{Possible discussion}
%\hl{yes, a table with existing signal processing  algorithms, their pluses and limitations is needed here}

\begin{table*}[htbp]
\caption{A comparison of different interference management techniques for detection (D), classification (C), localization (L), and mitigation (M) tasks.}
\begin{tabular}{p{2cm} p{6cm} p{4cm} p{4cm}}
\rowcolor{gray!25}  \textbf{Class} & \textbf{Approach and algorithms} & \textbf{Strengths} & \textbf{Limitations} \\
\hline
\rowcolor{gray!7}
Antenna based techniques (\modi{D, M, L}) & Use of antenna array to filter out interference signals, for example using angle of arrival. & Can handle different types of interferences. Outstanding performance. & It requires multiple receiver antennas, thus it has high hardware complexity. Multi-antenna receivers are sometimes subject to export control. \\
\rowcolor{gray!25}
Front-end techniques (\modi{D, M, L}) & Process signals before the analog to digital converter (ADC), e.g., Automatic Gain Control (AGC).  & Can handle different types of interferences. & Poor performance in the presence of low power level interference such as spoofing. \\
\rowcolor{gray!7}
Pre-correlation (\modi{D, C, M, L}) & Uses raw received signal parameters, e.g., frequency-domain or time-domain features (e.g., spectrogram, kurtosis, etc.); it can be based on both signal processing techniques, e.g. time power detector and probability distribution function detection, and ML-based techniques (e.g., RF fingerprinting).  & It can provide an early detection of interference presence and it can classify the interference types, typically using ML models; can reach large detection probabilities as the pre-correlation signal is not yet affected by many post-processing filtering stages. & It  has very limited capability to mitigate or localize the interferences and cannot pinpoint the exact satellites or pseudo-random (PRN) codes affected by interferences, it may require large training databases if ML is used. \\
\rowcolor{gray!25}
Post-correlation (\modi{D, C, M, L}) & Use signal   after correlation steps to detect inference, e.g., signal quality monitoring (SQM), power distortion monitoring (PDM), carrier-to-noise density ratio ($C/N_0$) monitoring, tracking peak monitoring, post-tracking scatter diagrams, etc. & Can detect, mitigate, and localize the interference; interference classification is typically not used at this stage; can identify each satellite whether genuine or spoofer.  & Inability to handle induced spoofing attacks due to dynamic distortions in correlation tracking loops; may also need large training databases if ML is used. \\
\rowcolor{gray!7}
 Navigation Level (\modi{D,M}) & Use multiple signals from multiple satellites, frequencies and receivers if available, e.g., angle-of-arrival discrimination, sum-of-squares for fault detection and exclusion via statistical consistency testing, jumps in clock and Doppler estimates, etc.  & %Likely to give the best results in interference mitigation compared to other receiver stages via statistical modeling, advanced estimation techniques and sensor fusion.  
 Integration with other sensors and positioning technologies is intuitive and well understood at this stage.
 & Long processing times as the full receiver chain needs to be completed before the navigation-level data is extracted as well as potential delays in interference response. \\
\end{tabular}
    \label{signalProcessingTable}
\end{table*}

% \LM{SUMMARY TABLE}
% \subsection{Classification}
% \textcolor{blue}{It is important to note that for efficient interference management, classification can come in hanbdy. In the literature, some works that focus on  classification include }

% \subsection{Mitigation}
% \textcolor{blue}{Mitigation techniques are focused on reducing (where possible eliminating) the impact of interference. }

% \subsection{Localization}
% \textcolor{blue}{The receiver can locate the source of interference..... }

%################
% \section{The role of AI in interference management}\label{sec:ai_role}
%\textcolor{red}{The most holistic approach to interference management in satellite is by relying on mechanisms that pereform all the. We envision that na feature extraction mechanism can come in handy in achieving the classification of different interferences. }

%################
\section{Vision for interference management }\label{sec:vision}

 Current GNSS receivers and associated solutions are a blend of classical signal processing and modern-day tools. Interference management is expected to continue evolving through innovative solutions that will exploit existing GNSS receiver architectures and advances in other areas such as wireless communications, space technologies, or AI/ML \cite{10379436}. Specifically, the use of AI/ML in GNSS is quickly growing, as in many other disciplines. Works such as \cite{9937069} provide an overview of the use of ML in the context of GNSS interference monitoring. The work highlights the superior performance of ML GNSS receiver algorithms compared to their conventional counterparts. The tutorial article \cite{10516283} also provides another valuable overview on the use of ML methodologies in various challenges associated with GNSS processing, particularly in urban environments. Another sign of growing interest in this area is \cite{closas2024emerging}, which summarizes articles in a recent special issue and provides a vision and outlook on future trends in the use of AI/ML for GNSS.

\begin{figure*}[t]
    \centering    \includegraphics[width=0.7\textwidth]{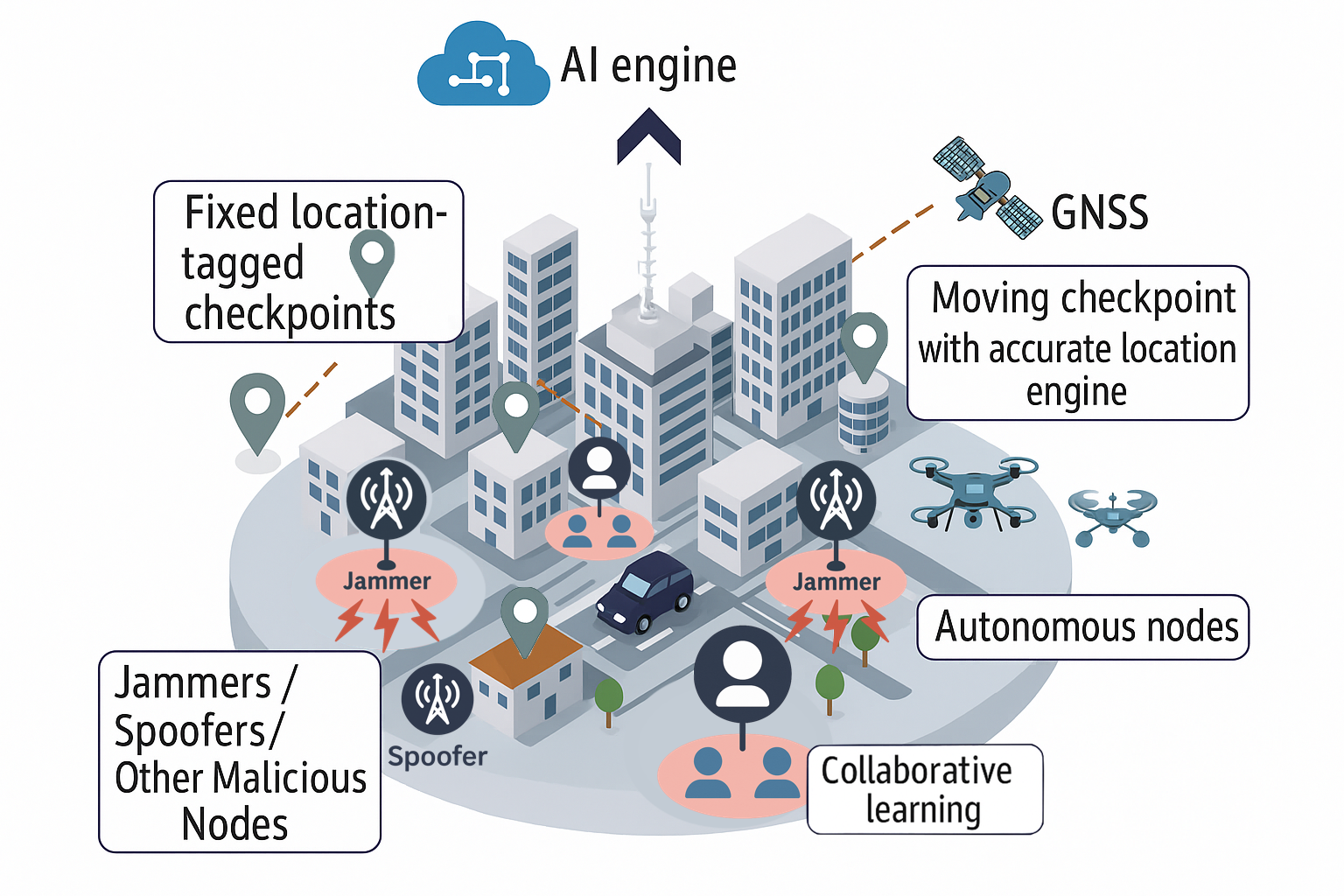}
    \caption{A scenario of moving users performing self-localization using GNSS and collaboratively enhancing interference management tasks, along with terrestrial- and space-borne infrastructure. Jamming and spoofing threats challenge navigation by denying/degrading it or deceiving the computated locations, respectively.}
    \label{fig:AI_Scenario}
\end{figure*}

In the context of interference monitoring, we can find \modi{the AI/ML} use in the four main \modi{elements} of detection, classification, mitigation and localization, \modi{but not necessarily independent of each other, but in an collaborative manner}. This section presents our vision and research outlook, aiming to anticipate where this exciting area is headed in the coming years. Fig.~\ref{fig:AI_Scenario} depicts the scenarios that we envision, where a multitude of connected agents interact with the environment and among themselves (in a centralized or decentralized manner) to navigate and contribute to interference management tasks. Additionally, those agents leverage existing infrastructure that is either terrestrial- or space-borne, as we discuss next. 

% Detailed paragraphs
\textbf{AI-aided interference detection \modi{and} classification.}
In detection and classification approaches, ML has received a lot of attention. For instance, Merwe \textit{et al.} in \cite{s23073452} introduces a \modi{low-cost} hardware GNSS-monitoring receiver that uses ML for both detection and classification. Their work showcase superior performance of this receiver compared to conventional receivers. Furthermore, Mehr \textit{et al.} in \cite{10681618} proposed an ML framework that uses  convolutional neural networks (CNN) for classification of jammers and other interferences using time frequency features. The results evince a performance of approximately $99\%$ in classifying interference. Chen \textit{et al.} in \cite{9760100} proposed a deep neural network \modi{(DNN)} for classifying interferences using spectrum fingerprint. While not constrained to feature-based classification, the proposed method achieved an average classification accuracy of $95\%$ in nine different interferers. Arguably, most existing GNSS interference detection methods require very large datasets, which are often impractical to collect, although proposals to leverage federated learning schemes have been recently proposed \cite{wu2025federated}. Alternatively, future AI approaches must efficiently leverage limited data to enhance both detection and classification performance. Another challenge with existing solutions is their inability to generalize to unseen interference types. This limitation causes models to perform accurate detection and classification only on data that resembles the data used during training. To address these two challenges, we envision that the few-shot learning and meta-learning frameworks detailed in \cite{10413635} can be highly beneficial. This is primarily because \modi{the} meta-learning frameworks optimize models to rapidly adapt to new interference scenarios using only a small amount of unseen data, thereby enabling fast fine-tuning of pretrained model, which is a crucial capability for next-generation GNSS receivers. \modi{Meta-learning approaches can also increase the generalisability of the detection algorithms, which is another desirable feature to cope with unknown interference types.}

Another potential but less explored solution is the inherent correlation of the signal structure across different receiver stages. We expect \modi{the} future GNSS interference detection and classification methods to leverage this correlation, either fully or partially. For instance, \modi{the} interference detection and classification ML models \modi{should use} features from both the pre-correlation and post-correlation \modi{receiver stages. This is done to enable access to different levels of data filtering, based on the hypothesis that data closer to the antenna preserves the transmitter hardware effects more effectively, through what is known as Radio Frequency Fingerprinting (RFF)\cite{soltanieh2020review}. Note that such RFF techniques are also more susceptible to noise compared to data obtained after the correlation stages.} \modi{As such, RFF-based solutions aim to} capture the unique imperfections of the transmitting device \modi{via ML}. This is because, even when devices are designed and manufactured according to the same standards, each one exhibits its own imperfections, which create a distinctive RFF signature \cite{9063411}. These imperfections affect the signal throughout all stages of the receiver, thereby inducing features that can be leveraged for classification and interference detection tasks in GNSS receivers.   

%\textcolor{blue}{Arguably, the most prominent ML tools that have been utilized involve vast training datasets for unsupervised learning approaches to learn impact of spoofing \cite{borhani2024detecting} and multipath \cite{li2022deep} on GNSS correlation, or their use in jamming classification \cite{MoralesFerre2019,10379436}; online learning schemes to learn hybrid data/physics-based models, a concept that is formalized in the augmented physics-based models (APBMs) framework presented in \cite{ImbiribaTAES23} to learn complex receiver dynamics or jamming localization in \cite{Nardin2023icassp}; the use of factor graph optimization (FGO), which has been shown to outperform existing solutions \cite{wen2021factor}; the use of federated learning (FL) to bring a layer of privacy in the necessary data sharing when data-driven models are trained, for instance in the context of jamming classifiers \cite{Peng2023plans}.}

\textbf{AI-aided interference mitigation.}
% mitigation: borio, rim, dnncorr
The use of AI for interference mitigation is \modi{currently} less explored \modi{than AI-based detection and classification} and has found \modi{little} interest in the community thus far. Some works investigate the use of deep learning models to filter out the effects of interferences \cite{mosavi2016narrowband,wei2016gps,Feiqiang25} or multipath \cite{li2022deep}. However, most of the past and ongoing efforts in terms of jamming mitigation resort to model-based approaches such as notch filtering and pulse blanking \cite{borio2016impact}; their generalizations based on robust statistics that form the corpus of solutions under the so-called robust interference mitigation (RIM) \cite{borio17ignss}; or the use of adaptive array processing \cite{fernandez2016robust}, which requires complex hardware designs that might not be feasible for mass-market deployments and typically remain exclusive of security or defense contexts.
That being said, other works in interference suppression using deep learning can be found in the general area of wireless communications, such as the survey \cite{oyedare2022interference}, from which GNSS research and development efforts could benefit from. \modi{Our vision blends the detection, classification and mitigation parts in a single broader task to be tackled with AI, where AI can play a trans formative role that adapts the mitigation method to the type of the detected interference and can even predict the future interference patterns. Solutions such as reinforcement learning (RL) and generative adversarial networks (GAN) \cite{Iqbal2024} seem well suited in this direction. The reason being that RL has the ability to dynamically adjust to the environment, while GAN has the ability to augment the training datasets in the case of scarce data, such as for rare interference types.}

\textbf{AI-aided interference localization.}
%\pau{add some more generic discussion here}
ML can be leveraged in crowdsourced environments to predict the received signal strength (RSS)\modi{-based interference} field and estimate the \modi{interferer}’s position \modi{for strong interferers such as jammers}~\cite{Jaramillo2025,Gattis2025,Jones2025plans}. In dense urban areas, where multipath and shadowing effects make localization particularly challenging, crowdsourced RSS measurements enable robust inference by leveraging data from multiple users. Work \cite{Jaramillo2025}  presents a comprehensive study on this approach, demonstrating how federated learning (FL) ensures data privacy while not degrading localization accuracy. By aggregating distributed observations without centralizing user data, FL improves scalability, reduces latency, and optimizes resource usage. This framework can be extended to also improve interference \modi{detection, classification, and} mitigation strategies, benefiting real-world GNSS security applications.

\begin{figure*}[t]
    \centering    \includegraphics[width=\textwidth]{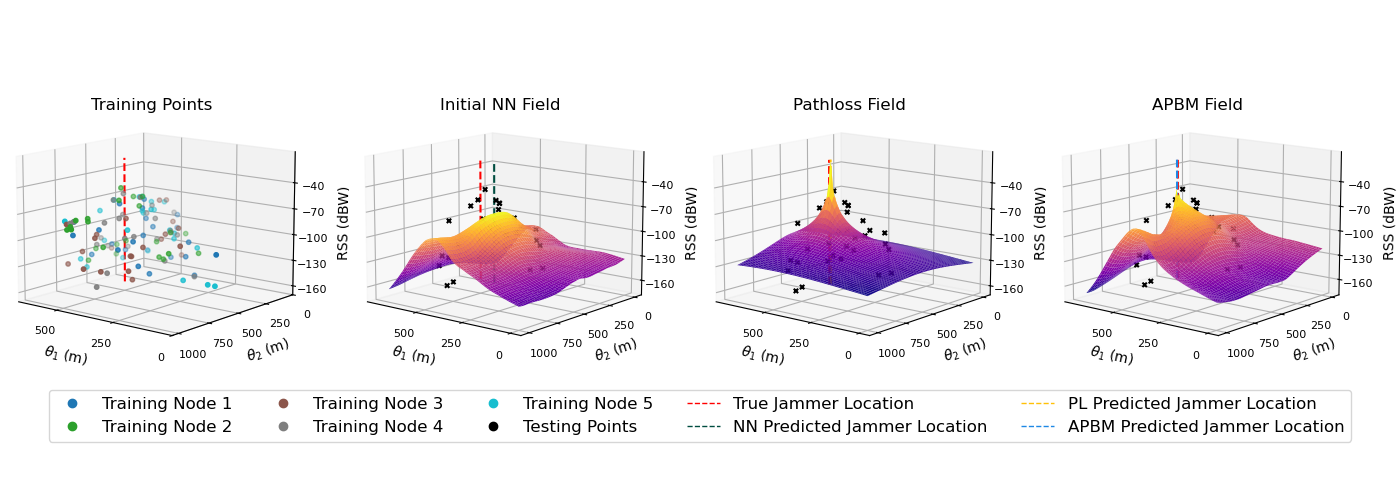}
    \caption{Jammer field prediction and localization in a urban ray-tracing scenario. From left to right: 
1) the training points distributed across multiple nodes used for model training, 
2) the predicted NN field with the NN-predicted jammer location, 
3) the predicted PL field with the PL-predicted jammer location, and 
4) the predicted APBM field with the APBM-predicted jammer location. The true jammer location is indicated for reference in each subplot.}
    \label{fig:localization_urban}
\end{figure*}

Fig.~\ref{fig:localization_urban} illustrates the \modi{envisioned} process. The first panel shows the training points across multiple clients. The second panel depicts a purely data-driven neural network (NN) approach, which estimates the RSS field but struggles to capture high RSS peaks due to the absence of a physics-based model. Despite this, the NN still provides a reasonable initial estimate of the jammer’s location (green marker) near the true location (red marker). Next, a path-loss (PL) model is used to model the propagation environment without ML. While it peaks at the jammer’s true location, it fails to capture high RSS values along major streets due to the absence of data-driven correction terms. Finally, the Augmented Physics-Based Model (APBM) introduced in \cite{Nardin2023icassp} fuses both NN and PL models, improving field prediction by incorporating physics-based constraints while leveraging data-driven corrections. The APBM produces a detailed field that accurately captures RSS variations and provides a refined jammer localization estimate. \modi{In view of this, the APBM framework presents potential for future research on jammer localization.}

\textbf{Distributed systems.} In an increasingly connected world, heterogeneous devices can effectively communicate among each other or through a central unit on the edge or cloud. Many works already exploit such pervasive deployment of devices and their seamless connectivity to share relevant information for interference management. We expect more works following this trend \modi{also in the GNSS field}, particularly in the context of 
$i)$ distributed processing (i.e. devices processing local data \modi{such as PNT data or geo-tagged data} to achieve common goals), where the challenge in some situations is to preserve \modi{the privacy} of information from the collaborative users (i.e. some users might be willing to collaborate but not at the expenses of sharing their location or personal information); 
$ii)$ swarms \modi{of devices, such as drones or tactical vehicles}, that is coordinated entities that share a common mission (e.g. navigate from one point to a destination, track an object, or survey an area maximizing its coverage). Those entities, which could be autonomous and ground or aerial borne, can automate \modi{the } interference monitoring tasks requiring minimal to no human interaction, thus reducing potential risks \cite{10542356}. \modi{AI role in here is to train the interference management models with collaborative data exchanged between the distributed agents in a privacy-preserving fashion and to send back to each agent parameter updates that allow the improvement of the local estimates.}

\textbf{Coexistence with heterogeneous \modi{multi-rate} systems.}
Interference management using technologies other than GNSS is currently being considered. \modi{These technologies are likely to operate at different rates with different physical layer specifications.} One of the most promising technical solutions is the use of LEO satellites \modi{either via data fusion with GNSS or} as \modi{LEO} sensing networks that can pinpoint the location of a GNSS jamming source. The main \modi{expected} features are high accuracy \modi{in interference detection and classification} (as LEO satellites are relatively close to Earth where interferences are), \modi{increased} resilience (\modi{LEO} satellites do not necessarily use GNSS \modi{on-board} to navigate \modi{and they may cope better than GNSS with interference through a higher frequency diversity}), and large coverage
\cite{Prol2022}.
Beyond the promise of LEO-based solutions \cite{kassas2024ad}, a variety of complementary technologies are available for which data fusion methods are required \cite{dardari2015indoor,dunik2020state}. \modi{Traditional data fusion approaches can be replaced with AI-based/data-driven fusion approaches in order to adapt better to complex environments and uncertainties. Also, AI could enable  dynamical fall-back mechanisms: when GNSS signals are highly affected by interference, the positioning solution can be obtained via alternative methods, such as via opportunistic LEO signals.}

%\ref{sec:int_threats}
\textbf{Adaptability to new threats.}
Beyond the known threats described in \cite{amin2016vulnerabilities}, GNSS interference monitoring system need to be aware of the appearance of new threats.  Those new attack modalities might \modi{take, for example, the form of} new jamming modulation schemes, more complex and potentially undetectable. For instance, systematic jamming \cite{curran17ignss,curran2016threat} or fuzzing \cite{yun2022fuzzing} pose new challenges that would require innovative solutions. A systematic jammer would target disrupting the correct reception of specific navigation bits, thus denying PNT service with a reduced jamming power when compared to standard jamming devices, making it nearly undetectable. \modi{Another example of new threats are in the form of }attacks that target complementary sensors, which are subject to their own vulnerabilities. While fusing GNSS \cite{dunik2020state} with other sensors (e.g. vision-based or inertial navigation) increases the defense firewalls, these also present opportunities for attackers to interfere with the overall PNT receiver. For instance, it was shown in \cite{narain2019security} that GPS/INS integration can be effectively spoofed under certain assumptions and attacker capabilities. The aforementioned trend to increase connectivity among devices and their cooperation also creates \modi{more} vulnerability since now attackers may leverage those aspects to generate new attack vectors. For instance, \cite{ranganathan2023analyzing,ranganathan2024impact} discuss the impact of GNSS spoofing on swarms of autonomous systems and how targeting the entire swarm or individual entities in the swarm can yield to different outcomes. \modi{As mentioned before, emerging AI paradigms such as meta-learning hold a promise to also adapt faster than the classical methods to the new threats.} \par

\section{Jammertest 2024 experiments}\label{Sec:jammertest}
\begin{figure}[htbp]
    \centering {\includegraphics[width=\columnwidth]{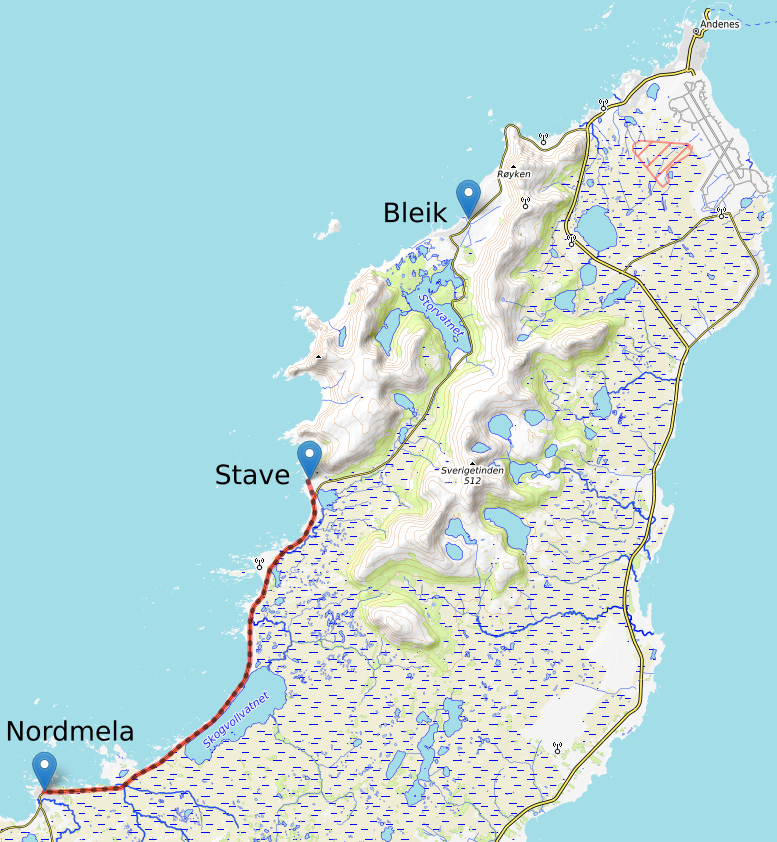}}
    \caption{The map of the Jammertest measurement locations. We collected data in Bleik and in the road between Stave and Nordmela, marked as a red dotted line. (Map: OpenStreetMap openstreetmap.org/copyright, basemap: OpenTopoMap)}
    \label{fig:andoya_map}
\end{figure}

% \pau{Introduce here Jammetest data, experiments, etc. to motivate the above.}
\modi{This section explores the potential of common benchmarks of wireless interference in GNSS bands for the research community and introduces the open dataset collected for further algorithm developoment.}
Due to legal restrictions \modi{which forbid the wireless transmission of interference in GNSS bands}, the lack of field measurement data significantly hinders resilient \modi{GNSS}-based positioning algorithm development. 
Therefore, we participated in the Jammertest 2024 event \modi{to collect and share a dataset of various jamming and spoofing incidents. Jammertest is the} largest field test event on GNSS \modi{wireless} jamming, spoofing, and meaconing worldwide, with open participation and no restrictions on data sharing or publication of results. Jammertest 2024 provided a test event in Bleik in Norway, including intentional jamming, spoofing, and meaconing of the GNSS reception in the Bleik village and roads nearby as shown in Fig.~\ref{fig:andoya_map}. The event attracted over $200$ participants from research institutions and companies and from $19$ countries worldwide \cite{Jammertest2024Underway2024,morrisonJammertest2022Jamming2023, broumandanHexagonNovAtelsJamming2024}. Jammertest is organized by the Norwegian Public Roads Administration, the Norwegian Communications Authority, the Norwegian Defense Research Establishment, the Norwegian Metrology Service and the Norwegian Space Agency. 

\begin{figure}[htbp]
    \centering {\includegraphics[width=\columnwidth]{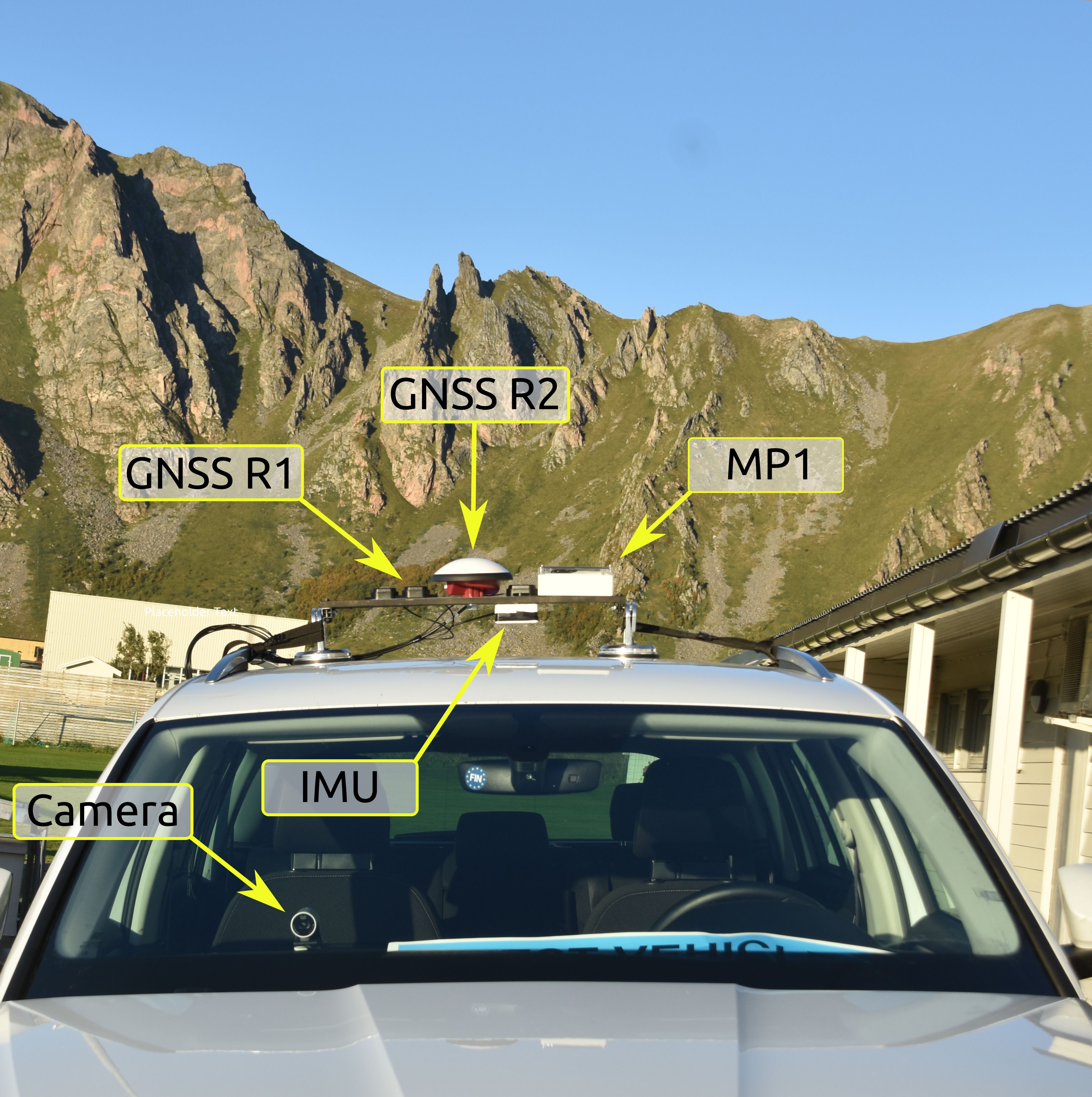}}
    \caption{The measurement setup in Bleik. The sensors are mounted on the roof of the car, while the jamming and spoofing devices are positioned on the mountaintop behind.}
    \label{fig:sensors}
\end{figure}

We collected data with a GNSS receiver. The following part shows some examples of data collected using affordable dual-band COTS GNSS receiver, later called as GNSS R1. The sensor installation is shown in Fig.~\ref{fig:sensors}. \modi{In addition to GNSS R1, the data collection also included another GNSS receiver, a mobile phone, an IMU, and a dashboard camera. The data is shared in GitHub in \url{https://github.com/UniVaasaDigiEco/VAARA} and further described in \cite{yliahoDataDrivenAnalysisPNT2025}}.

\begin{figure}[tbp]
    \centering {\includegraphics[width=\columnwidth]{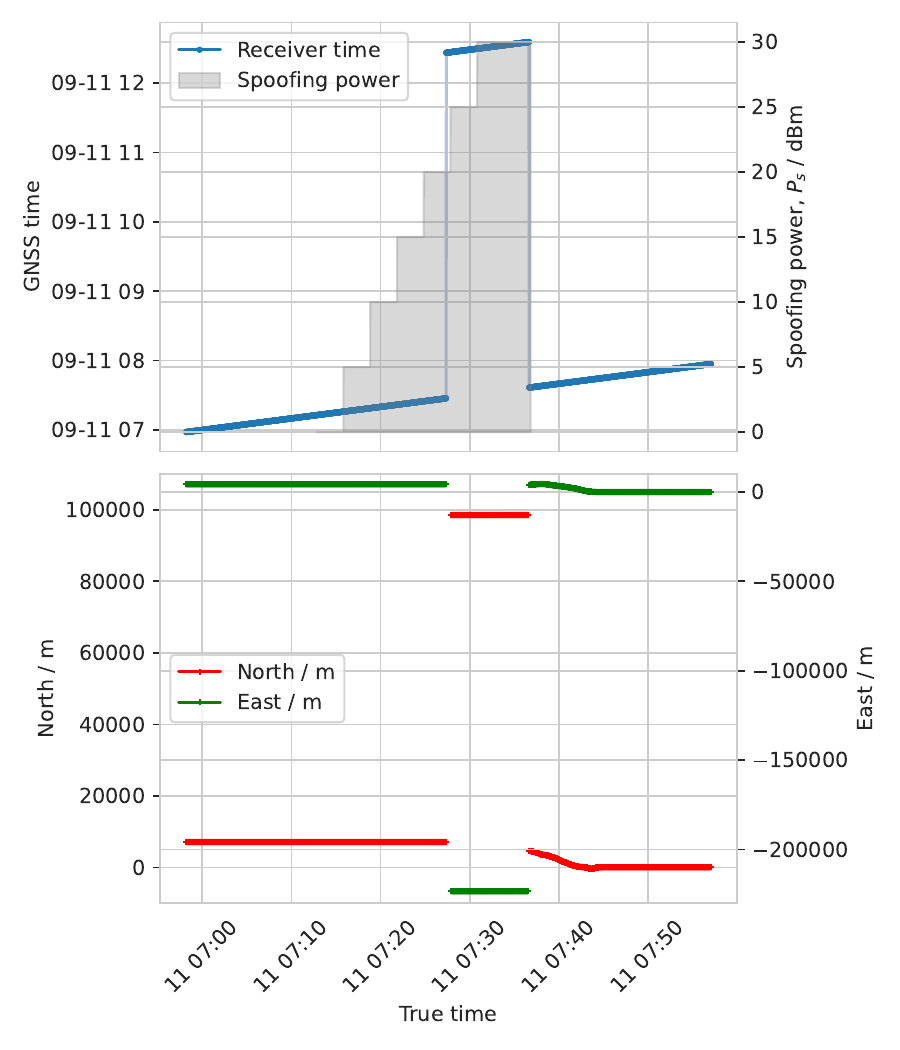}}
    \caption{Example of observed time and location spoofing during Jammertest. Spoofing power over 20 dBm started to affect the receiver causing both time and location solutions to jump at the same time.}
    \label{spoofsample}
\end{figure}

Fig.~\ref{spoofsample} illustrates an example of observed spoofing with GNSS R1. The spoofing started at 7:12:50 with power levels increasing in 5~dBm increments, each lasting three minutes. Once the power exceeded 20~dBm, it began to interfere with the receiver, resulting in a time shift of 5 hours and a location shift of more than 250~km. The receiver's operation returned to normal quickly after the spoofing event ended. This test allowed studying the \modi{spoofing} impact threshold and the recovery properties of the receiver. 

\begin{figure}[htbp]
	\centering {\includegraphics[width=\columnwidth]{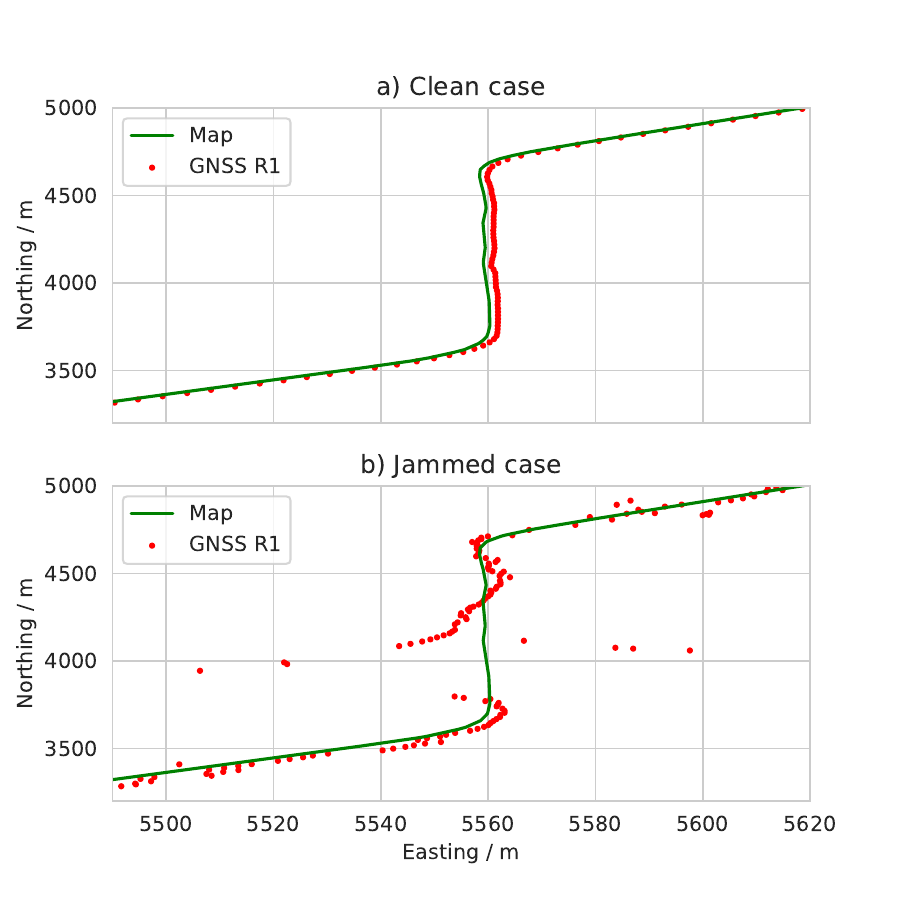}}
	\caption{Visualization of positioning accuracy with a COTS GNSS receiver (GNSS R1) in a clean case and jammed case. The map and GNSS R1 position estimates  are shown as green, and red colors respectively.}
	\label{spoofeffect}
\end{figure}

We also investigated the impact of jamming on positioning accuracy with the GNSS receiver. The jammer device was a handheld multi-frequency jammer, covering GNSS bands L1, L2 and L5. Its jamming power was between 1-1.6~W. Fig.~\ref{spoofeffect} shows a visualization of the receivers' performance, in comparison with a map as a benchmark. The accuracy of GNSS R1 receiver is substantially degraded in the presence of jamming interferences. \modi{While these results are as expected (i.e., jamming and spoofing do degrade the performance of GNSS receivers), the availability of such data rich in features, containing both spoofing and jamming in GNSS bands over wireless channels, is something to be brought to the attention of the research community. Currently, this data is available on demand and will soon be published as open access. Unlike the prior Jammertest campaigns before $2024$, the last year data provided a comprehensive and rich set of interference scenarios, which can become realistic benchmarks for testing the GNSS interference management solutions with various AI tools. }

\section{Conclusions and Outlook}\label{sec:conclusions}
In this paper, we presented a perspective on current and future GNSS interference management solutions \modi{and addressed the expected role of AI in GNSS interference management in the future}. We first provided an overview of existing approaches to interference management and highlighted their shortcomings, mainly due to their inability to handle complex interference scenarios, which are becoming increasingly common, \modi{as well as to generalize to unseen data}. Furthermore, we proposed ideas that can serve as a roadmap for future GNSS interference management. Specifically, we discussed the potential of leveraging AI models that can adapt through techniques such as meta-learning and few shot learning. Moreover, we \modi{highlighted} the need to develop methods that exploit \modi{the signal} correlations across different stages of the GNSS receiver as well as complementary sensors, which we believe have the potential to enhance the robustness of GNSS receivers. We also presented preliminary studies on leveraging a federated-learning framework to localize jammers \modi{in a distributed manner, by }using threat probability maps, thereby making threat localization more adaptive. Finally, the paper \modi{was} complemented with a set of real-data experiments conducted in the context of Jammertest 2024, where the impact of jamming and spoofing attacks \modi{was} highlighted \modi{and the necessity of having such benchmarks with realistic wireless interference was emphasized}. In summary, \modi{we believe that our} work provides a comprehensive resource for researchers working on AI-enabled GNSS.    
%\textcolor{blue}In this paper, we presented an overview of the current interference threats in global navigation satellite systems, which are on the rise due to the increased demand for precise time and navigation (PNT) services. Additionally, we reviewed state-of-the-art solutions and proposed potential AI-inspired approaches for managing interference, including detection, classification, localization and mitigation. For example, we discussed federated learning strategies aimed at collaboratively creating threat maps while preserving user identities. Ultimately, this information can be used to enhance navigation through message-passing algorithms that exploit the generated data.   
{}

\section*{Acknowledgment}
This work has been supported by the National Science Foundation under Awards ECCS-1845833 and CCF-2326559 and by the Research Council of Finland number 359846 (RESILIENT project). The authors used some AI tools to assist with minor language editing of this manuscript. AI tools were not used for content generation or analysis, but only to assist in generating the diagram in Fig.~\ref{fig:AI_Scenario}.

%The opinions expressed herein reflect the authors' view only. 

\balance
\bibliographystyle{ieeetr}
%\bibliography{output.bbl}
\bibliography{BibFile_max15items,closas-ref,petris-ref,leatileRef}

\end{document}